\documentclass[10pt, letterpaper, conference]{IEEEtran}
\usepackage[utf8]{inputenc}
\usepackage{graphicx}
\usepackage[export]{adjustbox}
\usepackage[labelfont=bf]{caption}
\usepackage{subcaption}
\usepackage{xcolor}
\usepackage{amsmath}
\usepackage{amssymb}
\usepackage{mathtools}
\usepackage[flushleft]{threeparttable}
\usepackage{pdfpages}

\usepackage{listings}
\usepackage{color}
\usepackage{multirow}

\definecolor{mygreen}{rgb}{0,0.6,0}
\definecolor{mygray}{rgb}{0.5,0.5,0.5}
\definecolor{mymauve}{rgb}{0.58,0,0.82}
\definecolor{myred}{RGB}{179,21,21}
\definecolor{myblue}{RGB}{34,18,181}

\title{BrainTTA: A 35 fJ/op Compiler Programmable Mixed-Precision Transport-Triggered NN SoC}
\author{\IEEEauthorblockN{M.J.~Molendijk\IEEEauthorrefmark{1}, 
    F.A.M.~de~Putter\IEEEauthorrefmark{1},
    M.~Gomony\IEEEauthorrefmark{1},
    P.~Jääskeläinen\IEEEauthorrefmark{4},
    H.~Corporaal\IEEEauthorrefmark{1}}
    \IEEEauthorblockA{\IEEEauthorrefmark{1}Electrical Engineering department, Eindhoven University of Technology, the Netherlands\\
    \texttt{\small \{m.j.molendijk, f.a.m.d.putter, m.gomony, h.corporaal\}@tue.nl}}
    \IEEEauthorblockA{\IEEEauthorrefmark{4}
Faculty of Information Technology and Communication Sciences, Tampere University, Finland\\}{\texttt{\small pekka.jaaskelainen@tuni.fi}}
    }

\begin{document}
\bstctlcite{BSTcontrol}

\maketitle

\begin{abstract}
    Recently, accelerators for extremely quantized deep neural network~(DNN) inference with operand widths as low as 1-bit have gained popularity due to their ability to largely cut down energy cost per inference. In this paper, a flexible SoC with mixed-precision support is presented. Contrary to the current trend of fixed-datapath accelerators, this architecture makes use of a flexible datapath based on a Transport-Triggered Architecture~(TTA). The architecture is fully programmable using C. The accelerator has a peak energy efficiency of 35/67/405~fJ/op (binary, ternary, and 8-bit precision) and a throughput of 614/307/77~GOPS, which is unprecedented for a programmable architecture.
\end{abstract}

\section{Introduction}
Edge computing is a rising computing paradigm with the ability to overcome privacy, latency, and energy issues that are currently being faced in the deployment of neural networks~(NNs) on embedded devices. While modern neural networks can solve complex tasks in the fields such as Computer Vision (CV) and Natural Language Processing (NLP), the sheer size of these networks prevents deploying such a network directly on embedded devices, which typically have limited storage capacity. Alongside the required storage for the parameters of these networks, the compute power also interferes with the deployment of such networks due to the energy constraints typically imposed on embedded hardware. To overcome these challenges, several approaches to optimize the models for low-power hardware have been explored. These include Hardware-aware Neural Architecture Search (NAS)~\cite{Tan2019Mnasnet:Mobile}, model compression in the form of pruning~\cite{Blalock2020WhatPruning} and quantization~\cite{Gholami2021AInference} and efficient data reuse~\cite{Waeijen2021ConvFusion:Networks}.

In parallel, research has been performed on creating highly specialized accelerators for neural network inference, exploiting the aforementioned model compression techniques. While these architectures perform great in terms of energy efficiency, the datapath structure is often not flexible and programmability is limited to some assembly dialect if programmable at all.

In short, current efforts towards low-power accelerators lack the flexibility to efficiently support different layers with varying sizes and varying parameter precision. Furthermore, the usability of these accelerators is hindered due to the absence of a compiler. In this paper, we present BrainTTA, the first flexible-datapath mixed-precision high-level programmable NN accelerator with compiler support. We showcase the flexibility of this architecture and the energy trade-off when running layers with different bit-widths. The contributions of this paper are threefold:
\begin{itemize}
    \item An energy-efficient TTA-based SoC for neural network inference supporting multiple precisions (binary, ternary and 8-bit) with a high energy efficiency of 35/67/405 fJ/op respectively (section~\ref{sec:arch_overview}~\&~\ref{sec:app_mapping}).
    
    \item  A thorough analysis of the system energy consumption for various operand bit-widths; it is found that the total system energy cost per operation grows superlinear with the bit-width of the operands~(section~\ref{sec:eval}).
    
    \item A detailed analysis on the trade-off between energy efficiency and flexibility. BrainTTA is more flexible than state-of-the-art architectures while limiting the overhead~(section~\ref{sec:related_work}). 
    
\end{itemize}

The remainder of this paper is organized as follows. In section~\ref{sec:backgr}, the background knowledge is discussed. Thereafter, in section~\ref{sec:arch_overview}, an overview of the full system architecture is discussed after which in section~\ref{sec:app_mapping}, the mapping of the network onto the proposed architecture is described. The results are presented in section~\ref{sec:eval} and a comparison with respect to state-of-the-art architectures is presented in section~\ref{sec:related_work}. The paper is concluded in section~\ref{sec:conclusion}.

\section{Background information}
\label{sec:backgr}
To relieve the burden on the memory and reduce the arithmetic complexity, quantization can be applied. Quantization can be applied down to 8-bit without significant loss of accuracy~\cite{Jacob2018QuantizationInference}. But even with 8-bit quantization, the storage requirements of modern networks are not in line with the storage size typically found in embedded hardware. Therefore, a push towards even lower bit-width quantization was made.

\subsection{Binary and Ternary Quantization}
Binary quantization restricts the weights and activations to a single bit; therefore the weights and activations are $w,a~\in~\{-1, +1\}$ whereas ternary \textit{trits} can additionally represent zero. This low operand precision introduces several advantages: the memory footprint is drastically reduced, the computations can be simplified, and the required bandwidth decreases sharply~\cite{Rastegari2016XNOR-net:Networks}\cite{Deng2018GXNOR-Net:Framework}. When both weights and activations are binarized or ternarized, the computations can be simplified by replacing the \texttt{MAC} (Multiply-Accumulate) operation with \texttt{XNOR} and \texttt{popcount} for binary and \texttt{Gated-XNOR} and \texttt{popcount} operations for ternary omitting the need for expensive multiplication operations.
\begin{figure}[t]
    \centering
    \includegraphics[width=0.9\linewidth]{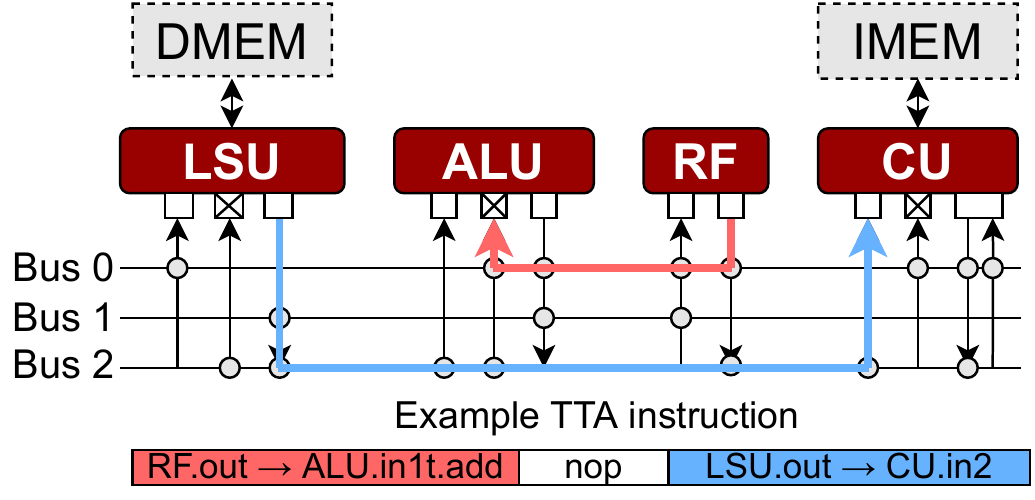}
    \caption{An example TTA instance and instruction, the square blocks denote \textit{input-} and \textit{output-ports}. A cross denotes a \textit{trigger-port}. The colored arrows drawn on the architecture illustrate the \textit{moves} inside the example instruction.}
    \label{fig:ttaBasic}
\end{figure}
\begin{figure*}[t]
     \centering
     \begin{subfigure}[b]{0.49\textwidth}
         \centering
         \includegraphics[width=\textwidth]{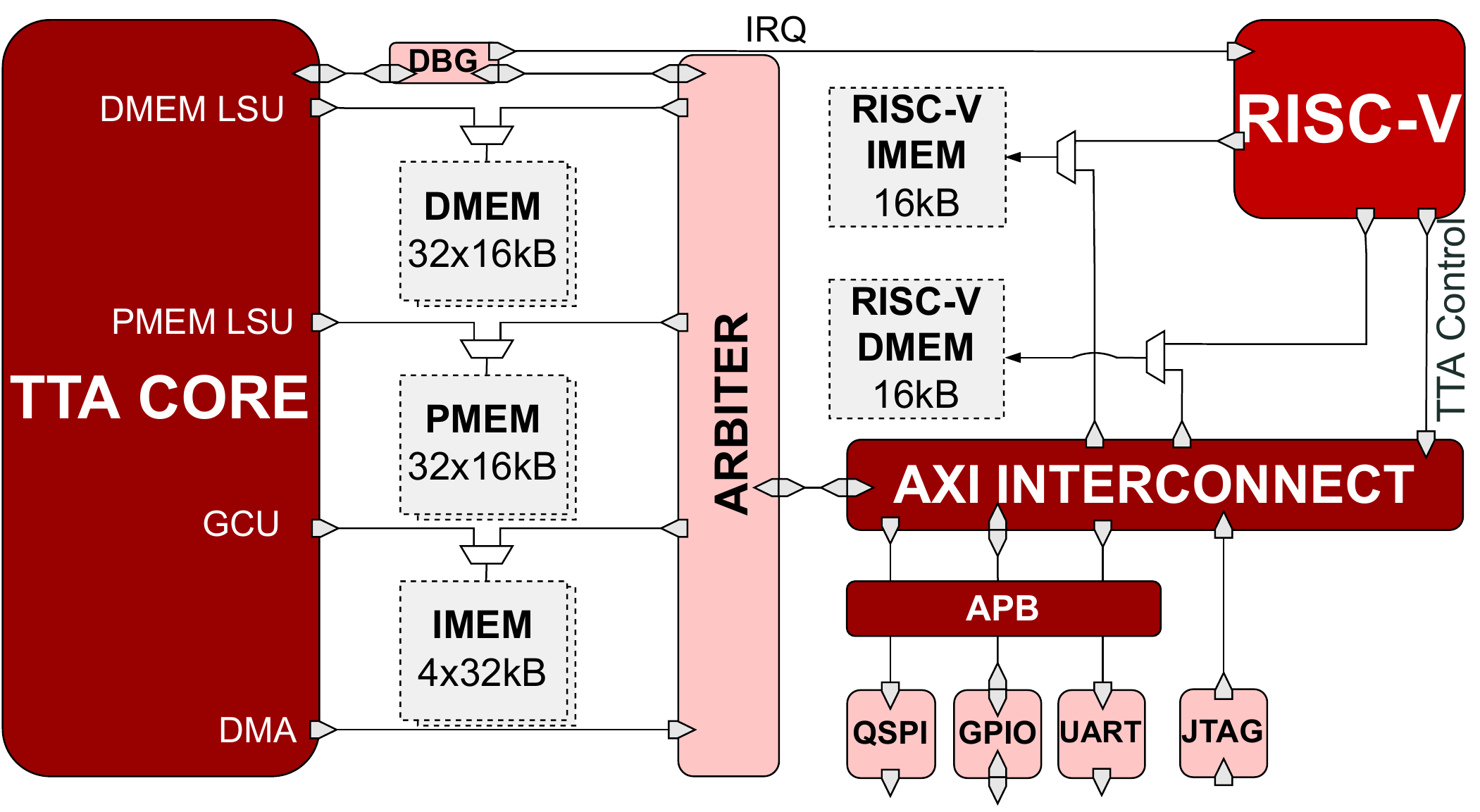}
         \caption{Block diagram of the BrainTTA SoC, the arbiter forms the border between the RISC and TTA part of the SoC.}
         \label{fig:ttaArchitecture}
     \end{subfigure}
     \hfill
     \begin{subfigure}[b]{0.45\textwidth}
         \centering
         \includegraphics[width=\textwidth]{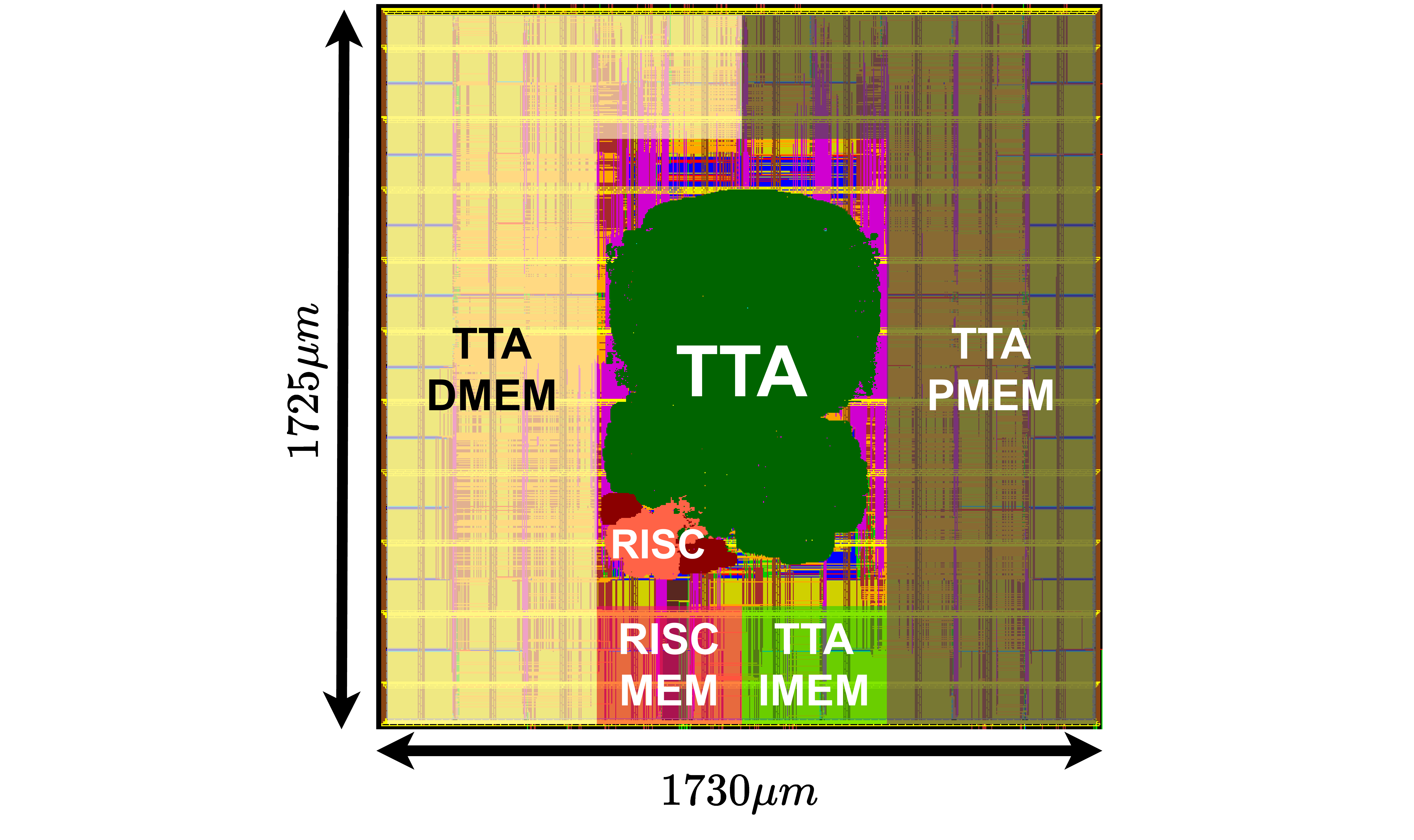}
         \caption{Layout (excluding pads), RISC logic (3.2\%) and TTA logic (28.0\%) are highlighted.}
         \label{fig:layout}
     \end{subfigure}
        \caption{Overview of the BrainTTA SoC.}
        \label{fig:BrainTTA_SoC}
\end{figure*}

While these extreme forms of quantization can offer significant energy and area savings, there is no such thing as a free lunch. There is still a significant gap in accuracy between 8-bit quantized networks and binary/ternary variants~\cite{Gholami2021AInference}\cite{Bulat2020XNOR-Net++:Networks}. Furthermore, some layers are more resilient to quantization than others~\cite{Gluska2020ExploringAnalysis}. This varying quantization penalty motivates the use of an architecture that supports mixed-precision, i.e. where different layers have different weight and activation bit-width.
\subsection{Transport-Triggered Architecture}
\label{sec:tta}
Transport-Triggered Architectures (TTA)~\cite{Corporaal1997MicroprocessorTTA} are programmed by specifying data movements instead of arithmetic operations, as typically found in VLIW architectures. This means that the movement of data is exposed to the compiler; the TTA is an explicit-datapath architecture, contrary to VLIW architectures, where data movement is implicit. With the compiler being in control of the data movements, optimizations like operand sharing and register file bypass can be exploited.

A basic instance of a TTA is displayed in Fig.~\ref{fig:ttaBasic}. The TTA contains a Control Unit (CU) used for instruction fetching and decoding, Register Files (RFs) for temporary storage, and Load-Store Units (LSUs) to access the memories. The connectivity (grey circles) is design-time configurable and can be made as generic or specific for certain applications as desired; more connectivity is at the expense of larger instruction size and more switching activity in the interconnect. In~\cite{Multanen2021Energy-EfficientProcessors}, several ways to reduce the instruction overhead, such as instruction compression are presented.

\section{Architectural overview}
\label{sec:arch_overview}

\begin{figure*}[t]
    \centering
    \includegraphics[width=1.0\linewidth]{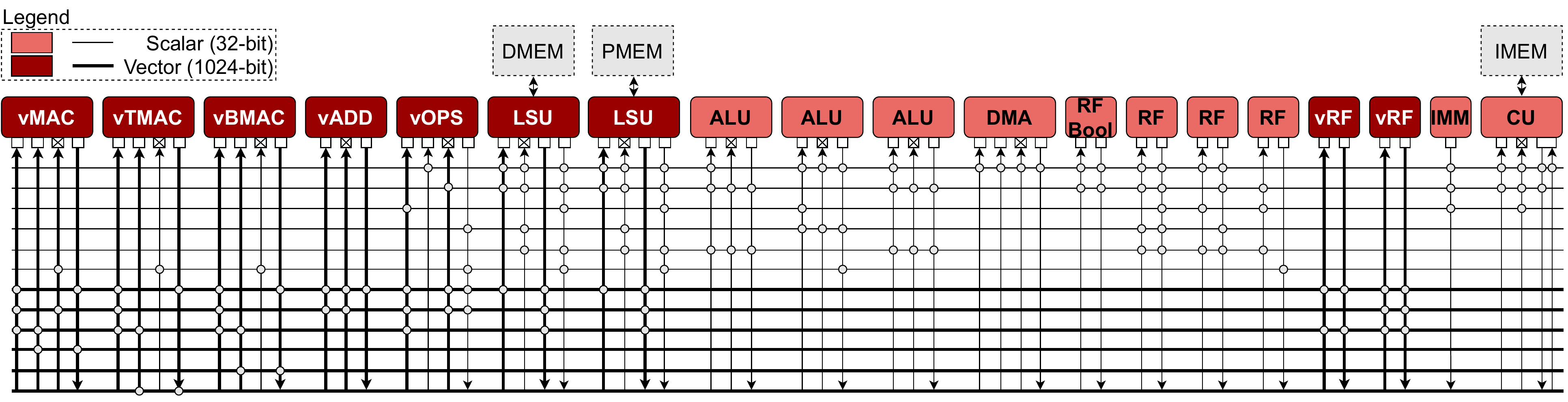}
    \caption{BrainTTA core instance, thicker lines denote vector buses, thinner lines scalar buses.}
    \label{fig:braintta}
\end{figure*}

The proposed full system architecture is displayed in Fig.~\ref{fig:BrainTTA_SoC} and the TTA core in Fig.~\ref{fig:braintta}. In this paper, the TTA-based Co-design Environment (TCE)~\cite{Jaaskelainen2016HW/SWProcessors} is used to create the TTA instance. This is an open-source toolchain that provides full compiler support. The compiler supports C, C++ and OpenCL. The TCE framework allows creation of any Functional Unit (FU) with arbitrary functionality and number of input and output ports. The FUs are controlled with custom instructions that can be added into the TCE compiler; the instructions can be inferred by the compiler or directly called using intrinsics. The main system components are:
\begin{itemize}
    \item \textit{RISC-V host processor}~\cite{Traber2017PULPino:Datasheet}, to start/stop the execution on the TTA core, initialize the memories and send and receive information via the external interfaces.
    \item \textit{TTA core}, used to perform the mixed-precision inference, more details will follow in the next paragraph.
    \item \textit{SRAM}, with separate memories for the RISC and TTA core, the TTA core. Memories are banked to allow efficient access of smaller bit-widths.
    \item \textit{Debugger (DBG)}, can halt the execution on the TTA core and signal completion of a task to the RISC-V host.
    \item \textit{AXI interconnect}, used for on-chip communication between the RISC and TTA-core and interfaces with the peripherals (APB) for off-chip communication. 
\end{itemize}

At the heart of the SoC is the TTA core. This core is used for the neural network inference and is based on TTA explained in Section~\ref{sec:tta}. The instantiation of the TTA core used in this paper is shown in Fig.~\ref{fig:braintta}. The core contains different FUs. The FUs are interconnected via buses, with scalar buses (bus 0-5) and vector buses (bus 6-11). The data transports (moves) on these buses are explicitly programmed. The core consists of the following units:

\textbf{Control Unit (CU)}; this unit contains the logic to fetch and decode instructions and steers the other units to execute the correct operations. Furthermore, the CU contains a hardware loopbuffer. Since the network layers are essentially described by multiple nested loops (listing~\ref{lst:naive_cnn}), having a hardware loopbuffer can greatly cut-down instruction fetch costs.

\textbf{Vector Multiply-Accumulate (vMAC/vBMAC/vTMAC)} unit is the workhorse of the TTA core. \texttt{MAC} operations are performed for binary, ternary, and 8-bit operands. The unit multiplies two 1024-bit vectors with 32 entries of 32-bits each, thus the vMAC contains 32 reduction trees, where each tree has 4 8-bit, 16 ternary, or 32 binary inputs. Input data reuse is exploited by broadcasting the input feature map to multiple units with different weights. The fixed vector width implies that for each bit-width, different vectorization factors are applied; this is further explained in Section~\ref{sec:vectorization}.

\textbf{Vector Add (vADD)} is used to add two (either 512- or 1024-bit) vectors, this can be used to support residual layers.

\textbf{Vector Operations (vOPS)}, auxiliary (vector) operations that are required in the network, alongside the computations. This FU can perform \texttt{quantization} as well as apply activation functions e.g. \texttt{ReLU} and pooling functions such as \texttt{MaxPool}. This FU also supports scalar element insertion and extraction on vectors.

\textbf{(Vector) Register Files (vRFs/RFs)} to store intermediate values or buffer weights to increase data reuse.

\textbf{Load-Store Units (LSUs)} are the interface between to the SRAM. There are two LSUs, mainly used to load weights (from PMEM) and one to load and store feature maps (DMEM). Since the memories are banked, multiples of 32-bit data can be loaded at low cost by selectively turning on memory banks.

\textbf{Scalar Units} are primarily used for address calculations needed as inputs to the LSUs.

The above mixture of units and supported operations is carefully selected after inspecting and scheduling frequently used DNN workloads.

\section{Application mapping} 
\label{sec:app_mapping}
The different layers in a neural network can generally be described in terms of nested for-loops, see listing~\ref{lst:naive_cnn} for an example. 

\newcommand\Comment{\hfill\normalfont\itshape}
\begin{lstlisting}[language=python, caption={An example of an 8-bit convolutional layer with output-stationary schedule, where $v_C$ and $v_M$ are the vectorization factors for the input- and output-channels.}, numbers=none, label=lst:naive_cnn, frame=tb]
for h in [0, H - R + 1]: @:\Comment \textcolor{myred}{Output feature map \textbf{height}} :@
  for w in [0, W - R + 1]: @:\Comment \textcolor{myred}{Output feature map \textbf{width}} :@
    for tm in [0, M/32]: @:\Comment \textcolor{myred}{\textbf{Ouput channels} ($v_M=32$)} :@
      acc = bias[tm]
      for tc in [0, C/4]: @:\Comment \textcolor{myred}{\textbf{Input channels} ($v_C=4$)} :@
        for r in [0, R]: @:\Comment \textcolor{myred}{Kernel \textbf{height}} :@
          for s in [0, S]: @:\Comment \textcolor{myred}{Kernel \textbf{width}} :@
            acc += in_buffer[h + r][w + s][c] * 
                   weights[n][r][s][tm]
      out_buffer[h][w][tm] = acc
\end{lstlisting}

Since applications for the TTA can be programmed in C, the schedule can easily be altered for each layer separately. Changing the schedule simply boils down to applying loop transformations (e.g. unroll, interchange, tile). This scheduling freedom, in combination with the exposed-datapath operating principle, allows the creation of an efficient schedule on a per-layer basis with minimized data movements.
\subsection{Layer Support}
Different neural networks constitute different layer types. Among the layer types that are supported in BrainTTA are:
\begin{enumerate}
    \item Convolutional layer \hfill (8b in, 32b out)
    \item Binary convolutional layer \hfill (1b in, 16b out)
    \item Ternary convolutional layer \hfill (2b in, 16b out)
    \item Depth-wise convolutional layer \hfill (8b in, 32b out)
    \item Fully-connected layer \hfill(8b in, 32b out)
    \item Residual addition \hfill(16/32b in, 16/32b out)
    \item Requantization \hfill(to 8b, 2b or 1b)
\end{enumerate}
All above layers are supported by BrainTTA, an energy breakdown of the first three is given in Section~\ref{sec:eval}.

The supported layers are described by:

\textbf{Convolutional layers} are supported with three different bit-widths, namely 8-bit, ternary and binary. Depending on the bit-width of the convolutional layer, the 1024-bit weight and input vector are split in different ways (more information in section~\ref{sec:vectorization}). Since different output feature maps use the same input feature maps, input broadcasting is possible for data reuse.

\textbf{Depth-wise convolutional layers} are supported by changing the scalar-vector product used in convolutional layers to vector-vector products which is required since each weight kernel is bound to a single input channel; in other words, input broadcasting is not possible. 

\textbf{Fully-connected layers} execution is similar to that of convolutional layers, however, the kernel size is now 1x1.

\textbf{Residual addition} is adding two higher bit-width values, but to support this, the scaling factor of the values added together needs to match. The latter is called \textbf{requantization}. 

To keep down the overhead that comes with the flexibility and programmability of BrainTTA, parallelism is introduced in several scheduling dimensions (i.e. dimensions that are shown in listing~\ref{lst:naive_cnn}); in the next paragraph, the choice of vectorization dimensions to achieve this parallelism will be elaborated.

\subsection{Vectorization}
\label{sec:vectorization}
\begin{figure}[t]
    \centering
    \includegraphics[width=1.0\linewidth]{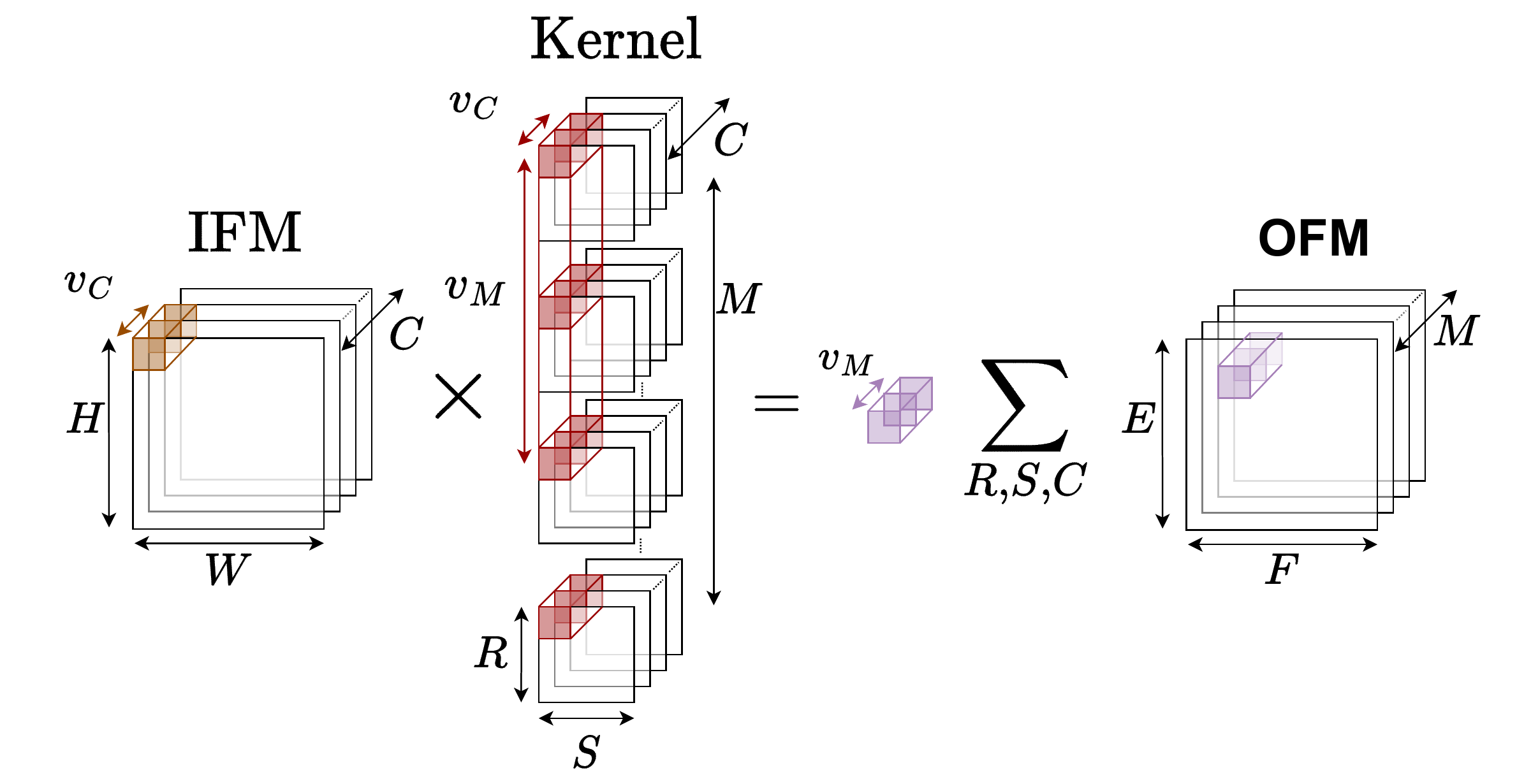}
    \caption{A convolutional layer where the Input Feature Map~(IFM), Kernel and Output Feature Map~(OFM) vectorization is visualized; $v_C$ is vectorization over the IFM channel dimension, $v_M$ over the OFM channel dimension.}
    \label{fig:vectorization}
\end{figure}

The vectorization is visualized in Fig.~\ref{fig:vectorization}. The choice of vectorization dimensions is based on three observations. First of all, modern networks typically have more feature maps (higher $C$, $M$) but the size of each individual feature map can be smaller ($W$, $H$)~\cite{He2016DeepRecognition}\cite{Szegedy2016Inception-v4Learning}, this means that to populate a large vector, one should not only vectorize over $W$ and $H$. Secondly, the \texttt{MAC}, binary and ternary \texttt{popcount} operations produce an intermediate output value with a much higher bit-width than the quantized value; requantization should happen as soon as possible to reduce movement of large intermediate values. Therefore, the final value of a single pixel in the output feature map should be calculated as early as possible, which makes an output-stationary schedule favorable. Lastly, the \texttt{popcount} outputs the sum of its inputs. Therefore, the inputs supplied to the popcount module should contribute to the same output pixel. This means that the inputs supplied to a single popcount module (which corresponds to a single vector element), should either have a different $W \text{\&} H$ indices in the same receptive field or from a different input channel $C$.  In BrainTTA, the datapath width is 1024-bits and the output vectorization factor is $v_M = 32$; therefore, to fill the 1024-bit datapath, $v_C = 32$ for binary, $v_C = 16$ for ternary and $v_C = 4$ for 8-bit inputs.

\section{Results}
\label{sec:eval}
The design that is shown in Fig.~\ref{fig:ttaArchitecture} is synthesized using GlobalFoundries 22nm FDX technology using an operating voltage of 0.5V while targeting the typical corner. After synthesis, the layout is created which can be seen in Fig.~\ref{fig:layout}. 

\subsection{Experimental setup}
The flow that is used to go from RTL to layout consists of Cadence Genus 21.10 for the logic synthesis and Cadence Innovus 21.11 for the back-end implementation. These tools are also used to obtain energy numbers. The energy numbers are obtained by annotating the switching activity found during the post-layout simulation in order to gain the most accurate energy figures possible.

\subsection{Post-layout simulation results}

\begin{figure*}[t]
     \centering
        \includegraphics[width=0.9\textwidth]{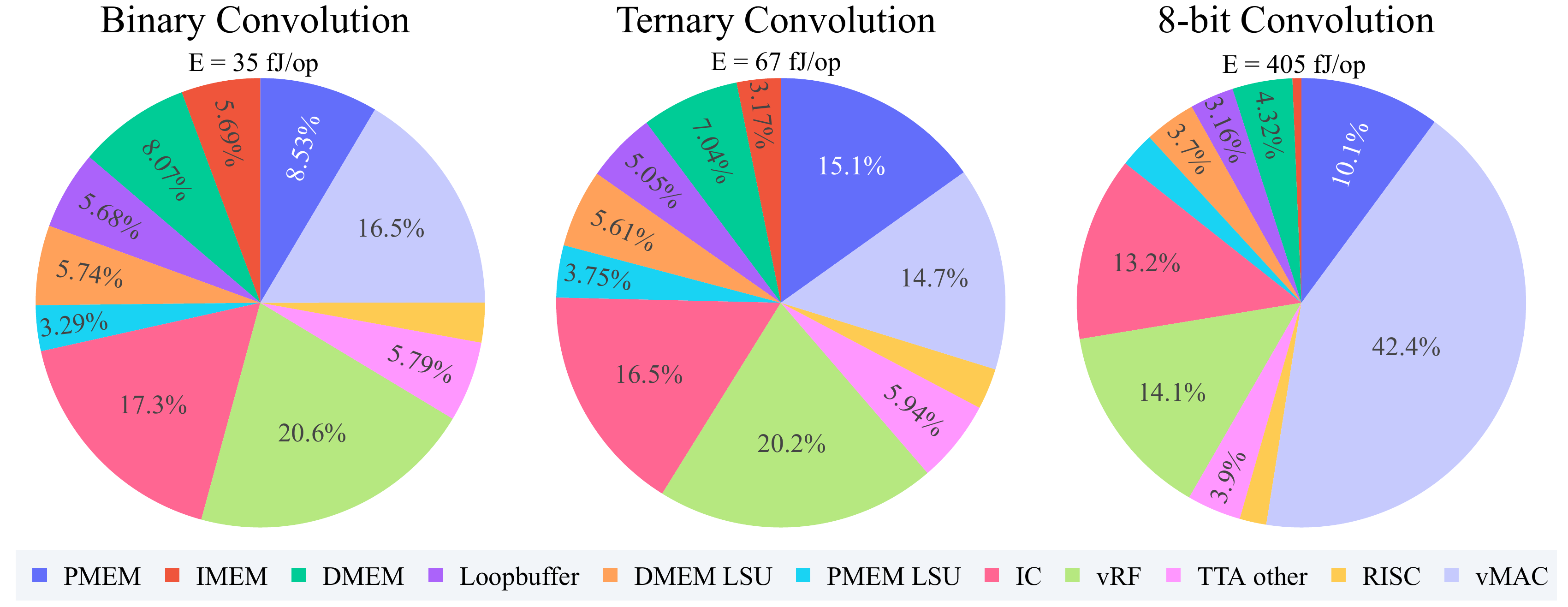}
        \caption{Energy breakdown for the convolutional layers (GF 22nm FDX, 300~MHz, 0.5V, R=S=3, M=C=128 and W=H=16).}
        \label{fig:piechart}
\end{figure*}
The area of the SoC is $3.0mm^2$ excluding IO pads as can be seen in Fig.~\ref{fig:layout}. The largest part of the floorplan is dedicated to the data (DMEM) and parameter (PMEM) memory of the TTA core, holding the input/output feature map and weights respectively. Both the DMEM and PMEM are made by combining 32 banks of 16kB each, resulting in a combined data storage capacity of 1MB.

Fig.~\ref{fig:piechart} displays the energy that is required to perform three convolution layers, with binary, ternary, and 8-bit operands. A \texttt{MAC} is counted as two operations. It can be seen that the energy per operation difference between the binary and ternary convolution is nearly a factor of 2. Furthermore, the breakdowns between the two different scenarios are very similar with the exception of the instruction memory. The reason for this similarity is that the compute unit (vMAC) circuitry and usage of the binary and ternary convolution are very alike, and their utilization of the other components is identical \textit{but} the amount of computations per second is halved since the ternary digits (trits) take up twice the space of single binary digits, hence the doubled energy per operation. In other words, the overhead per computation is doubled when doubling the number of bits of the operands.

Furthermore, the interconnect (IC) of the TTA core takes second place in energy usage in the logic after the vMAC. This is one of the architectural characteristic components where a price is paid for flexibility. In contrast, the fixed-datapath architectures discussed in the related work do not provide the freedom to freely move data around between different units in a programmable way. The routing flexibility in BrainTTA in combination with the freedom to implement any FU (and retain compiler support) makes it possible to run more complex networks like ResNet and even non-DNN workloads independently of the general purpose processor.

\section{Comparison with State-of-the-art}
\label{sec:related_work}
Various papers have been published about architectures containing accelerators to bring down the energy per inference as much as possible. Most of these accelerators can be grouped into: fully-digital implementations, mixed-signal approaches and Compute-in-Memory (CIM) approaches. Although CIM architectures~\cite{Valavi2019ACompute}\cite{Bankman2019AnCMOS} can currently go beyond 500 \mbox{TOPS/W}, they suffer from the design time required for the complex analog compute blocks. Furthermore, these analog compute blocks show chip-to-chip variation due to CMOS process variation which makes them error-prone and difficult to benchmark the exact performance. Therefore, the related work focuses on fully digital implementations, specifically designed for inference on heavily quantized networks.
\subsection{Related work}
XNOR-Neural Engine (XNE)~\cite{Conti2018XNORInference} and ChewBaccaNN~\cite{Andri2020ChewBaccaNN:Accelerator} are binary accelerators that support the previously described reduced arithmetic complexity introduced by binary quantization. In~\cite{Conti2018XNORInference}, a full SoC is presented able to run neural networks autonomously with the help of a configurable Micro-Controller Unit (MCU). Due to the lack of input data reuse, the SRAM reads dominate the energy consumption. This is circumvented by turning the SRAMs completely off and storing all data in RFs, therefore making it unable to run any realistic workloads. In~\cite{Andri2020ChewBaccaNN:Accelerator}, SRAM is avoided at all by using SCM only. Although the authors claim flexibility, the kernel size is hard-wired into the design (with a dramatic drop in performance when using a different kernel size).

In~\cite{Scherer2020CUTIE:Efficiency} a ternary accelerator called CUTIE is presented. By spatially unrolling all the convolutional loops (seen in Listing~\ref{lst:naive_cnn}), iterating can be avoided and great data reuse is guaranteed if the convolutional loop iterators match the hardware design point. Spatially unrolling the loop dimensions directly constrains the network layer sizes that can be run efficiently on the hardware, therefore sacrificing flexibility.

In~\cite{Cho2021ReconfigurableScheme}, the idea of having scheduling flexibility in BNN inference is explored. The schedule flexibility is motivated by the change in layer parameters from the first layers into the deeper layers. The authors report two different schedules implemented in hardware, both relying on feature map parallelism. One for the shallow layers ($w,h > c$) and one for the deeper layers ($c > w, h$). Although an increase in throughput is reported by adding scheduling flexibility, power numbers for ASIC implementation are omitted, therefore this paper is not taken into account in the actual comparison.

In~\cite{Knag2021ACMOS}, a binary accelerator is implemented based on the compute-near-memory principle. All kernel sizes are hard-wired, therefore it offers little to no flexibility.
\subsection{Comparison}
The BrainTTA is compared to state-of-the-art accelerators in Table~\ref{tab:comparison_sota}. It is partitioned into three different sets: the general properties of each architecture, the neural network layer requirements for full utilization of the chip, and the support of other layers and programmability. BrainTTA beats XNE in energy efficiency, while XNE is the only accelerator that has comparable flexibility to BrainTTA. This flexibility does incur overhead; to lower the overhead, the hardware loopbuffer was used to reduce instruction fetches and data reuse was implemented using a combination of data broadcasting and parameter buffering in the vector register files. CUTIE, ChewBaccaNN and \cite{Knag2021ACMOS} all have a higher efficiency per operation, however, these architectures are not programmable and flexibility of these architectures is severely lacking due to the hard-wired datapaths. Furthermore, if a look is taken at the dimensions that are hard-wired into the design of the SotA competitors, attaining the peak throughput and peak efficiency becomes very challenging since the neural network layers would need to adhere to very specific layer size requirements. Illustrative to this problem is ChewBaccaNN, which reports a maximum throughput of 240 GOPS while only achieving 23 GOPS in XNORNET-++~\cite{Bulat2020XNOR-Net++:Networks}. Furthermore, none of these architectures supports 8-bit operands or is programmable in a high-level language.

\begin{table*}[]
\centering
\caption{Comparison of performance, efficiency and flexibility of the architectures discussed.}
\label{tab:comparison_sota}
\begin{threeparttable}
\begin{tabular}{lrclcrrr}
 & \multicolumn{1}{c}{\textbf{ChewBaccaNN}~\cite{Andri2020ChewBaccaNN:Accelerator}} & \multicolumn{2}{c}{\textbf{CUTIE}~\cite{Scherer2020CUTIE:Efficiency}} & \multicolumn{2}{c}{\textbf{XNE}~\cite{Conti2018XNORInference}} & \multicolumn{1}{c}{\textbf{10nm FinFet}~\cite{Knag2021ACMOS}} & \multicolumn{1}{c}{\textbf{BrainTTA}} \\ \hline
\textbf{Implementation characteristics} & \multicolumn{1}{l}{} & \multicolumn{1}{l}{} &  & \multicolumn{1}{l}{} & \multicolumn{1}{l}{} & \multicolumn{1}{l}{} & \multicolumn{1}{l}{} \\
Technology node [nm] & 22 & \multicolumn{2}{c}{22} & \multicolumn{2}{c}{22} & 10 & 22 \\
Supply voltage [V] & 0.4 & \multicolumn{2}{c}{0.65} & \multicolumn{1}{r}{0.6} & 0.4 & 0.39 & 0.5 \\
Inference precision\tnote{1} & b & \multicolumn{2}{c}{b\tnote{2}, t} & \multicolumn{2}{c}{b} & b & b, t, i8 \\
Memory technology & SCM & \multicolumn{1}{r}{SRAM} & \multicolumn{1}{r}{SCM} & \multicolumn{1}{r}{SRAM} & SCM & SRAM & SRAM \\ \hline
\textbf{Key Performance Indicators} & \multicolumn{1}{l}{} & \multicolumn{1}{l}{} &  & \multicolumn{1}{l}{} & \multicolumn{1}{l}{} & \multicolumn{1}{l}{} & \multicolumn{1}{l}{} \\
Peak throughput [GOPS] & 240 & \multicolumn{2}{c}{16000} & \multicolumn{1}{r}{67} & 5 & 3400 & 614 \\
Energy/op [fJ] binary & 4.48/15.38\tnote{3} & \multicolumn{2}{c}{-} & \multicolumn{1}{r}{115} & 21.6 & 1.62 & 35 \\
Energy/op [fJ] ternary & - & \multicolumn{1}{r}{2.19} & \multicolumn{1}{r}{1.70} & \multicolumn{2}{c}{-} & - & 67 \\
Energy/op [fJ] 8-bit & - & \multicolumn{2}{c}{-} & \multicolumn{2}{c}{-} & - & 405 \\
Core area [mm$^2$] & 0.7 & \multicolumn{2}{c}{7.5} & \multicolumn{2}{c}{2.32} & 0.39 & 2.98 \\
Area efficiency [GOPS/mm$^2$] & 343 & \multicolumn{2}{c}{2133} & \multicolumn{2}{c}{28.88} & 8717 & 206 \\
Memory capacity [kB] (excl. instruction memory) & 153 & \multicolumn{1}{r}{-} & \multicolumn{1}{r}{-} & \multicolumn{1}{r}{520} & 16 & 161 & 1024 \\ \hline
\textbf{Flexibility} &  & \multicolumn{1}{r}{} &  & \multicolumn{1}{r}{} & \multicolumn{1}{l}{} &  &  \\
Full utilization for & \multicolumn{1}{l}{} & \multicolumn{1}{l}{} &  & \multicolumn{1}{l}{} & \multicolumn{1}{l}{} & \multicolumn{1}{l}{} & \multicolumn{1}{l}{} \\
\quad Number of IFMs (C) multiple of & 16 & \multicolumn{2}{c}{128} & \multicolumn{2}{c}{128} & 1024 & 32/16/4\tnote{4} \\
\quad Number of OFMs (M) multiple of & Any & \multicolumn{2}{c}{128} & \multicolumn{2}{c}{128} & 128 & 32 \\
\quad Kernel height (R) is & 7 & \multicolumn{2}{c}{3} & \multicolumn{2}{c}{Any} & 2 & Any \\
\quad Kernel width (S) is & 7 & \multicolumn{2}{c}{3} & \multicolumn{2}{c}{Any} & 2 & Any \\
Partial result support (for scheduling freedom) & Yes & \multicolumn{2}{c}{No} & \multicolumn{2}{c}{No} & No & Yes \\
Residual layer support & Yes & \multicolumn{2}{c}{No\tnote{5}} & \multicolumn{2}{c}{No} & No & Yes \\
Programmability & None & \multicolumn{2}{c}{None} & \multicolumn{2}{c}{None} & None & C/C++/OpenCL \\ \hline
\end{tabular}
\begin{tablenotes}
    \item[1] b = \texttt{binary}, t = \texttt{ternary}, i8 = \texttt{integer8}.
    \item[2] Only estimates were provided, under the assumption that all ternary-specific hardware is removed.
    \item[3] For 7x7 and 3x3 convolution respectively.
    \item[4] For \texttt{binary}, \texttt{ternary} and \texttt{integer8} respectively.
    \item[5] Partial result support is not needed since the output pixel computation is fully unrolled in hardware.
\end{tablenotes}
\end{threeparttable}
\end{table*}

\section{Conclusion}
\label{sec:conclusion}
A novel TTA-based SoC for neural network inference that seamlessly combines flexibility with efficiency is presented in this paper. The SoC is able to perform operations at 35/67/405~fJ/op for binary, ternary and 8-bit operands respectively. Still, it is highly flexible and can easily adapt to different types of networks such that it can advance together with the algorithmic inventions in the area of heavily quantized neural networks. The support for mixed-precision allows the SoC to mitigate accuracy loss in layers that are most adversely affected by low bit-width quantization (typically the first and last layer of the network). The mixed-precision and compiler support, in combination with the exposed-datapath to minimize data movement make this architecture very versatile while approaching the energy efficiency of much less flexible architectures of competitors. \textbf{Future work:} the authors still see options for future improvements such as including ternary compression, adding more vMAC units to sequentially calculate outputs in a systolic way, and intertwining the compiler with a hardware-aware mapping tool such as ZigZag~\cite{Mei2021ZigZag:Accelerators}.


\bibliographystyle{IEEEtran}
\bibliography{ref_control.bib,references.bib}

\end{document}